\documentclass[10pt,twocolumn,twoside]{IEEEtran}
\usepackage{graphicx}
\usepackage{amsmath}

\usepackage{amsfonts}
\usepackage{amssymb}
\usepackage{array}
\newcommand{\PreserveBackslash}[1]{\let\temp=\\#1\let\\=\temp}
\newcolumntype{C}[1]{>{\PreserveBackslash\centering}p{#1}}
\newcolumntype{R}[1]{>{\PreserveBackslash\raggedleft}p{#1}}
\newcolumntype{L}[1]{>{\PreserveBackslash\raggedright}p{#1}}
\usepackage[usenames]{color}
\usepackage{colortbl,booktabs}
\usepackage{tabularx}
\usepackage{multicol}
\usepackage{booktabs}
\usepackage[Symbol]{upgreek}
\usepackage{subfigure}
\usepackage{bm}
\usepackage{cite}
\usepackage{balance}
\usepackage[ruled,vlined]{algorithm2e}

\usepackage{threeparttable}
\usepackage{amsmath}
\usepackage{dcolumn}
\usepackage{multirow}
\usepackage{amsfonts}
\newcolumntype{d}[1]{D{.}{.}{#1}}

\begin{document}

\bibliographystyle{IEEEtran} 
\title{Channel Estimation for RIS Assisted Wireless Communications: Part II - An Improved Solution Based on Double-Structured Sparsity \\ {\large {(Invited Paper)}}}

\author{Xiuhong Wei, Decai Shen, and Linglong Dai
\vspace{-0.8em}
\thanks{All authors are with the Beijing National Research Center for Information Science and Technology (BNRist) as well as the Department of Electronic Engineering, Tsinghua University, Beijing 100084, China (e-mails: weixh19@mails.tsinghua.edu.cn, sdc18@mails.tsinghua.edu.cn, daill@tsinghua.edu.cn).}
\thanks{This work was supported in part by the National Key Research and Development Program of China (Grant No. 2020YFB1807201) and in part by the National Natural Science Foundation of China (Grant No. 62031019).}
}

\maketitle
\vspace{-2em}
\begin{abstract}
     Reconfigurable intelligent surface (RIS) can manipulate the wireless communication environment by controlling the coefficients of RIS elements. However, due to the large number of passive RIS elements without signal processing capability, channel estimation in RIS assisted wireless communication system requires high pilot overhead. In the second part of this invited paper, we propose to exploit the double-structured sparsity of the angular cascaded channels among users to reduce the pilot overhead. Specifically, we first reveal the double-structured sparsity, i.e., different angular cascaded channels for different users enjoy the completely common non-zero rows and the partially common non-zero columns. By exploiting this double-structured sparsity, we further propose the double-structured orthogonal matching pursuit (DS-OMP) algorithm, where the completely common non-zero rows and the partially common non-zero columns are jointly estimated for all users. Simulation results show that the pilot overhead required by the proposed scheme is lower than existing schemes.

\end{abstract}

\begin{IEEEkeywords}
Reconfigurable intelligent surface (RIS), channel estimation, compressive sensing.
\end{IEEEkeywords}

\section{Introduction}\label{S1}

In the first part of this two-part invited paper, we have introduced the fundamentals, solutions, and future opportunities of channel estimation in the reconfigurable intelligent surface (RIS) assisted wireless communication system. One of the most important challenges of channel estimation is that, the pilot overhead is high, since the RIS consists of a large number of passive elements without signal processing capability~\cite{Bichai, Renzo_RISvsRelay}. By exploiting the sparsity of the angular cascaded channel, i.e., the cascade of the channel from the user to the RIS and the channel from the RIS to the base station (BS), the channel estimation problem can be formulated as a sparse signal recovery problem, which can be solved by compressive sensing (CS) algorithms with reduced pilot overhead~\cite{JunCS,LiangCS}. However, the pilot overhead of most existing solutions is still high.

In the second part of this paper, in order to further reduce the pilot overhead, we propose a double-structured orthogonal matching pursuit (DS-OMP) based cascaded channel estimation scheme by leveraging the double-structured sparsity of the angular cascaded channels\footnote{Simulation codes are provided to reproduce the results presented in this paper: http://oa.ee.tsinghua.edu.cn/dailinglong/publications/publications.html.
}. Specifically, we reveal that the angular cascaded channels associated with different users enjoy the completely common non-zero rows and the partially common non-zero columns, which is called as ``double-structured sparsity" in this paper. Then, by exploiting this double-structured sparsity, we propose the DS-OMP algorithm based on the classical OMP algorithm to realize channel estimation. In the proposed DS-OMP algorithm, the completely common row support and the partially common column support for different users are jointly estimated, and the user-specific column supports for different users are individually estimated. After detecting all supports mentioned above, the least square (LS) algorithm can be utilized to obtain the estimated angular cascaded channels. Since the double-structured sparsity is exploited, the proposed DS-OMP based channel estimation scheme is able to further reduce the pilot overhead.

The rest of the paper is organized as follows. In Section II, we introduce the channel model and formulate the cascaded channel estimation problem. In Section III, we first reveal the double-structured sparsity of the angular cascaded channels, and then propose the DS-OMP based cascaded channel estimation scheme. Simulation results and conclusions are provided in Section IV and Section V, respectively.

{\it Notation}: Lower-case and upper-case boldface letters ${\bf{a}}$ and ${\bf{A}}$ denote a vector and a matrix, respectively; ${{{\bf{a}}^T}}$ denotes the conjugate of vector $\bf{a}$; ${{{\bf{A}}^T}}$ and ${{{\bf{A}}^{H}}}$ denote the transpose and conjugate transpose of matrix $\bf{A}$, respectively; ${{\left\|  \bf{A}  \right\|_F}}$ denotes the Frobenius norm of matrix ${\bf{A}}$; ${\rm{diag}}\left({\bf{x}}\right)$ denotes the diagonal matrix with the vector $\bf{x}$ on its diagonal; ${\bf{a}}{\otimes}{\bf{b}}$ denotes the Kronecker product of ${\bf{a}}$ and ${\bf{b}}$.
Finally, $\cal CN\left(\mu,\sigma \right)$ denotes the probability density function of the circularly symmetric complex Gaussian distribution with mean $\mu$ and variance $\sigma^2$.
\vspace{-3mm}

\section{System Model}\label{S2}
In this section, we will first introduce the cascaded channel in the RIS assisted communication system. Then, the cascaded channel estimation problem will be formulated.

\subsection{Cascaded Channel}\label{S2.1}
We consider that the BS and the RIS respectively employ the $M$-antenna and the $N$-element uniform planer array (UPA) to simultaneously serve ${K}$ single-antenna users. Let $\bf{G}$ of size ${M \times N}$ denote the channel from the RIS to the BS, and ${\bf{h}}_{r,k}$ of size ${N \times 1}$ denote the channel from the ${k}$th user to the RIS $\left({k = 1,2, \cdots ,K}\right)$. The widely used Saleh-Valenzuela channel model is adopted to represent ${\bf{G}}$ as~\cite{Hu18}
\begin{equation}\label{eq1}
{\bf{G}}=\sqrt{\frac{MN}{L_G}}\sum\limits_{l_1 = 1}^{L_G}{\alpha^{G}_{l_1}}{\bf{b}}\left( \vartheta^{G_r}_{l_1},\psi^{G_r}_{l_1}\right){{\bf{a}}\left( \vartheta^{G_t}_{l_1},\psi^{G_t}_{l_1}\right)}^{T},
\end{equation}
where $L_G$ represents the number of paths between the RIS and the BS, ${\alpha}^{G}_{l_1}$, $\vartheta^{G_r}_{l_1}$ (${\psi}^{G_r}_{l_1}$), and ${\vartheta}^{G_t}_{l_1}$ (${\psi}^{G_t}_{l_1}$) represent the complex gain consisting of path loss, the azimuth (elevation) angle at the BS, and the azimuth (elevation) angle at the RIS for the $l_1$th path. Similarly, the channel ${\bf{h}}_{r,k}$ can be represented by
\begin{equation}\label{eq2}
{\bf{h}}_{r,k}=\sqrt{\frac{N}{L_{r,k}}}\sum\limits_{l_2 = 1}^{L_{r,k}}{\alpha^{r,k}_{l_2}}{{\bf{a}}\left( \vartheta^{r,k}_{l_2},\psi^{r,k}_{l_2}\right)},
\end{equation}
where $L_{r,k}$ represents the number of paths between the $k$th user and the RIS, ${\alpha}^{r,k}_{l_2}$, $\vartheta^{r,k}_{l_2}$ (${\psi}^{r,k}_{l_2}$) represent the complex gain consisting of path loss, the azimuth (elevation) angle at the RIS for the $l_2$th path. ${\bf{b}}\left( \vartheta,\psi\right) \in {\mathbb{C}^{M\times 1}}$ and ${\bf{a}}\left( \vartheta,\psi\right) \in {\mathbb{C}^{N\times 1}}$ represent the normalized array steering vector associated to the BS and the RIS, respectively. For a typical $N_1\times N_2$ ($N=N_1\times N_2$) UPA, ${\bf{a}}\left(\vartheta,\psi\right)$ can be represented by~\cite{Hu18}
\begin{equation}\label{eq3}
{\bf{a}}\left(\vartheta,\psi \right) = \frac{1}{{\sqrt N }}{\left[ {{e^{ - j2{\pi}d{\rm{sin}}\left(\vartheta\right){\rm{cos}}\left(\psi\right) {\bf{n}}_1/{\lambda}}}} \right]}{\otimes}{\left[ {{e^{ - j2{\pi}d{\rm{sin}}\left(\psi\right) {\bf{n}}_2/{\lambda}}}} \right]},
\end{equation}
where ${\bf{n}}_1=[0,1,\cdots,N_1-1]$ and ${\bf{n}}_2=[0,1,\cdots,N_2-1]$, $\lambda$ is the carrier wavelength, and
$d$ is the antenna spacing usually satisfying $d = \lambda/2$.

Further, we denote ${\bf{H}}_{k}\triangleq{\bf{G}}{\rm{diag}}\left({\bf{h}}_{r,k}\right)$ as the $M\times N$ cascaded channel for the $k$th user. Using the virtual angular-domain representation, ${\bf{H}}_{k}\in\mathbb{C}^{M\times N}$ can be decomposed as
\begin{equation}\label{eq4}
{{\bf{H}}_{k}} = {\bf{U}}_{M}{\tilde{\bf{H}}}_{k}{\bf{U}}_{N}^{T},
\end{equation}
where ${\tilde{\bf{H}}}_{k}$ denotes the ${M \times N}$ angular cascaded channel, ${\bf{U}}_{M}$ and ${\bf{U}}_{N}$ are respectively the ${M \times M}$ and ${N \times N}$ dictionary unitary matrices at the BS and the RIS~\cite{Hu18}. Since there are limited scatters around the BS and the RIS, the angular cascaded channel ${\tilde{\bf{H}}}_{k}$ has a few non-zero elements, which exhibits the sparsity.

\subsection{Problem Formulation}\label{S2.2}
In this paper, we assume that the direct channel between the BS and the user is known for BS, which can be easily estimated as these in conventional wireless communication systems~\cite{Hu18}. Therefore, we only focus on the cascaded channel estimation problem.

By adopting the widely used orthogonal pilot transmission strategy, all users transmit the known pilot symbols to the BS via the RIS over ${Q}$ time slots for the uplink channel estimation. Specifically, in the ${q}$th $\left({q = 1,2, \cdots ,Q}\right)$ time slot, the effective received signal ${{\bf{y}}_{k,q}}\in\mathbb{C}^{M\times 1}$ at the BS for the $k$th user after removing the impact of the direct channel can be represented as
\begin{equation}\label{eq5}
\begin{aligned}
{{\bf{y}}_{k,q}} = & {\bf{G}}{\rm{diag}}\left({\bm{\theta}}_{q}\right){\bf{h}}_{r,k}{s}_{k,q}+{\bf{w}}_{k,q}
\\ = & {\bf{G}}{\rm{diag}}\left({\bf{h}}_{r,k}\right){\bm{\theta}}_{q}{s}_{k,q}+{\bf{w}}_{k,q},
\end{aligned}
\end{equation}
where $s_{k,q}$ is the pilot symbol sent by the $k$th user, ${\bm{\theta}}_{q}=[\theta_{q,1},\cdots,\theta_{q,N}]^T$ is the $N\times 1$ reflecting vector at the RIS with $\theta_{q,n}$ representing the reflecting coefficient at the $n$th RIS element $(n=1, \cdots,N)$ in the $q$th time slot, ${{\bf{w}}_{k,q}}\sim{\cal C}{\cal N}\left( {0,\sigma^2{\bf{I}}_{M}} \right)$ is the ${{M} \times 1}$ received noise with ${\sigma^2}$ representing the noise power. According to the cascaded channel ${\bf{H}}_{k}={\bf{G}}{\rm{diag}}\left({\bf{h}}_{r,k}\right)$, we can rewrite~(\ref{eq5}) as
\begin{equation}\label{eq56}
{{\bf{y}}_{k,q}}={\bf{H}}_{k}{\bm{\theta}}_q{s}_{k,q}+{\bf{w}}_{k,q}.
\end{equation}

After ${Q}$ time slots of pilot transmission, we can obtain the ${M \times Q}$ overall measurement matrix ${\bf{Y}}_k={[{\bf{y}}_{k,1}, \cdots, {\bf{y}}_{k,Q}]}$ by assuming ${s_{k,q}} = 1$ as
\begin{equation}\label{eq6}
{{\bf{Y}}_{k}}={\bf{H}}_{k}{\bm{\Theta}}+{\bf{W}}_{k},
\end{equation}
where ${\bm{\Theta}}={[{\bm{\theta}}_{1}, \cdots ,{\bm{\theta}}_{Q}]}$ and ${\bf{W}}_{k}=[{\bf{w}}_{k,1},\cdots,{\bf{w}}_{k,Q}]$.
By substituting~(\ref{eq4}) into~(\ref{eq6}), we can obtain
\begin{equation}\label{eq7}
{{\bf{Y}}_{k}}={\bf{U}}_{M}{\tilde{\bf{H}}}_{k}{\bf{U}}_{N}^{T}{\bm{\Theta}}+{\bf{W}}_{k}.
\end{equation}

Let denote ${{\tilde{\bf{Y}}}_{k}}=\left({\bf{U}}_{M}^{H}{{\bf{Y}}_{k}}\right)^{H}$ as the $Q\times M$ effective measurement matrix, and ${{\tilde{\bf{W}}}_{k}}=\left({\bf{U}}_{M}^{H}{{\bf{W}}_{k}}\right)^{H}$ as the $Q\times M$ effective noise matrix,~(\ref{eq6}) can be rewritten as a CS model:
\begin{equation}\label{eq8}
{{\tilde{\bf{Y}}}}_{k}={\tilde{\bm{\Theta}}}{\tilde{\bf{H}}}_{k}^{H}+{\tilde{\bf{W}}}_{k},
\end{equation}
where ${{\tilde{\bf{\Theta}}}}=\left({\bf{U}}_{N}^{T}{{\bm{\Theta}}}\right)^{H}$ is the $Q\times N$ sensing matrix. Based on~(\ref{eq8}), we can respectively estimate the angular cascaded channel for each user $k$ by conventional CS algorithms, such as OMP algorithm. However, under the premise of ensuring the estimation accuracy, the pilot overhead required by the conventional CS algorithms is still high.


\section{Joint Channel Estimation for RIS Assisted Wireless Communication Systems}\label{S3}
In this section, we will first reveal the double-structured sparsity of the angular cascaded channels. Then, by exploiting this important channel characteristic, we will propose a DS-OMP based cascaded channel estimation scheme to reduce the pilot overhead. Finally, the computational complexity of the proposed scheme will be analyzed.

\subsection{Double-Structured Sparsity of Angular Cascaded Channels}\label{S3.1}
In order to further explore the sparsity of the angular cascaded channel both in row and column, the angular cascaded channel ${\tilde{\bf{H}}}_{k}$ in~(\ref{eq4}) can be expressed as
\begin{equation}\label{eq9}
\begin{aligned}
{\tilde{\bf{H}}}_k=&\sqrt{\frac{MN}{L_GL_{r,k}}}\sum\limits_{l_1 = 1}^{L_G}\sum\limits_{l_2 = 1}^{L_{r,k}}{\alpha^{G}_{l_1}}{\alpha^{r,k}_{l_2}}\\&{\tilde{\bf{b}}}\left( \vartheta^{G_r}_{l_1},\psi^{G_r}_{l_1}\right){\tilde{\bf{a}}^{T}\left( \vartheta^{G_t}_{l_1}+\vartheta^{r,k}_{l_2},\psi^{G_t}_{l_1}+\psi^{r,k}_{l_2}\right)},
\end{aligned}
\end{equation}
where both ${\tilde{\bf{b}}}\left(\vartheta,\psi\right)= {\bf{U}}_{M}^{H}{\bf{b}}\left(\vartheta,\psi\right)$ and ${\tilde{\bf{a}}}\left( \vartheta,\psi\right)={\bf{U}}_{N}^{H}{{\bf{a}}}\left( \vartheta,\psi\right)$ have only one non-zero element, which lie on the position of array steering vector at the direction $\left( \vartheta,\psi\right)$ in ${\bf{U}}_{M}$ and ${\bf{U}}_{N}$. Based on~(\ref{eq9}), we can find that each complete reflecting path $(l_1, l_2)$ can provide one non-zero element for ${\tilde{\bf{H}}}_{k}$, whose row index depends on $\left(\vartheta^{G_r}_{l_1},\psi^{G_r}_{l_1}\right)$ and column index depends on $\left(\vartheta^{G_t}_{l_1}+\vartheta^{r,k}_{l_2},\psi^{G_t}_{l_1}+\psi^{r,k}_{l_2}\right)$. Therefore, ${\tilde{\bf{H}}}_{k}$ has $L_G$ non-zero rows, where each non-zero row has $L_{r,k}$ non-zero columns. The total number of non-zero elements is $L_GL_{r,k}$, which is usually much smaller than $MN$.

\begin{figure}[htbp]
\begin{center}
\includegraphics[width=0.7\linewidth]{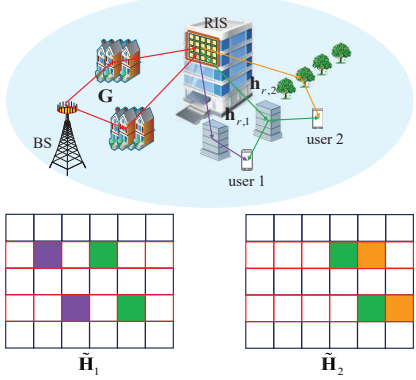}
\end{center}
\setlength{\abovecaptionskip}{-0.3cm}
\caption{Double-structured sparsity of the angular cascaded channels.} \label{RIS}
\end{figure}

More importantly, we can find that different sparse channels $\{{\tilde{\bf{H}}}_{k}\}_{k=1}^K$ exhibit the double-structured sparsity, as shown in Fig. 1. Firstly, since different users communicate with the BS via the common RIS, the channel $\bf{G}$ from the RIS to the BS is common for all users. From~(\ref{eq9}), we can also find that $\bigg\{\left( \vartheta^{G_r}_{l_1},\psi^{G_r}_{l_1}\right)\bigg\}_{l_1=1}^{L_G}$ is independent of the user index $k$. Therefore, the non-zero elements of $\{{\tilde{\bf{H}}}_{k}\}_{k=1}^K$ lie on the completely common $L_G$ rows. Secondly, since different users will share part of the scatters between the RIS and users, $\{{\bf{h}}_{r,k}\}_{k=1}^K$ may enjoy partially common paths with the same angles at the RIS. Let $L_c$ (${L_{c}}\leq{L_{r,k}}, \forall k$) denote the number of common paths for $\{{\bf{h}}_{r,k}\}_{k=1}^K$, then we can find that for $\forall l_1$, there always exists $\bigg\{\left( \vartheta^{G_t}_{l_1}-\vartheta^{r,k}_{l_2},\psi^{G_t}_{l_1}-\psi^{r,k}_{l_2}\right)\bigg\}_{l_2=1}^{L_c}$ shared by $\{{\tilde{\bf{H}}}_{k}\}_{k=1}^K$. That is to say, for each common non-zero rows $l_1$ ($l_1=1,2,\cdots,L_G$), $\{{\tilde{\bf{H}}}_{k}\}_{k=1}^K$ enjoy $L_c$ common non-zero columns. This double-structured sparsity of the angular cascaded channels can be summarized as follows from the perspective of row and column, respectively.
\begin{itemize}

\item Row-structured sparsity: Let $\Omega_r^{k}$ denote the row set of non-zero elements for ${\tilde{\bf{H}}}_{k}$, then we have
    \begin{equation}\label{16}
      \Omega_r^{1}=\Omega_r^{2}=\cdots=\Omega_r^{K}=\Omega_r,
    \end{equation}
    where $\Omega_r$ represents the completely common row support for $\{{\tilde{\bf{H}}}_{k}\}_{k=1}^K$.
\item Partially column-structured sparsity: Let $\Omega_c^{l,k}$ denote the column set of non-zero elements for the $l_1$th non-zero row of ${\tilde{\bf{H}}}_{k}$, then we have
    \begin{equation}\label{17}
      \Omega_c^{l_1,1}\cap\Omega_c^{l_1,2}\cap\cdots\cap\Omega_c^{l_1,K}=\Omega_c^{l_1,{\rm{Com}}},\quad l_1=1,2,\cdots,L_G,
    \end{equation}
    where $\Omega_c^{{l,{\rm{Com}}}}$ represents the partially common column support for the $l_1$th non-zero row of $\{{\tilde{\bf{H}}}_{k}\}_{k=1}^K$.
\end{itemize}

Based on the above double-structured sparsity, the cascaded channels for different users can be jointly estimated to improve the channel estimation accuracy.

\subsection{Proposed DS-OMP Based Cascaded Channel Estimation}\label{S3.2}

In this subsection, we propose the DS-OMP based cascaded channel estimation scheme by integrating the double-structured sparsity into the classical OMP algorithm. The specific algorithm can be summarized in \textbf{Algorithm 1}, which includes three key stages to detect supports of angular cascaded channels.

\begin{algorithm}[htbp]
\caption{DS-OMP based cascaded channel estimation}
\textbf{Input}: ${\tilde{\bf{Y}}}_k:\forall k$, ${\tilde{\bm{\Theta}}}$, ${L_G}$, ${L_{r,k}}:\forall k$, $L_c$.
\\\textbf{Initialization}: ${\hat{\tilde{\bf{H}}}}_k={\bf{0}}_{M\times N},\forall k$.
\\1. \textbf{Stage 1:} Return estimated completely common row support ${\hat{\Omega}}_r$ by \textbf{Algorithm 2}.
\\2. \textbf{Stage 2:} Return estimated partially common column supports $\{{\hat{\Omega}}_c^{l_1,{\rm{Com}}}\}_{l_1=1}^{L_G}$ based on ${\hat{\Omega}}_r$  by \textbf{Algorithm 3}.
\\3. \textbf{Stage 3:} Return estimated column supports $\{\{{\hat{\Omega}}_c^{l_1,k}\}_{l_1=1}^{L_G}\}_{k=1}^K$ based on ${\hat{\Omega}}_r$ and $\{{\hat{\Omega}}_c^{l_1,{\rm{Com}}}\}_{l_1=1}^{L_G}$ by \textbf{Algorithm 4}.
\\4. \textbf{for} $l_1 = 1,2,\cdots,L_G$ \textbf{do}
\\5. \hspace*{+3mm} \textbf{for} $k = 1,2,\cdots,K$ \textbf{do}
\\6. \hspace*{+6mm} ${\hat{\tilde{\bf{H}}}}^H_k({\hat{\Omega}}_c^{l_1,k},{{\hat{\Omega}}_{r}}(l_1))={\tilde{\bm{\Theta}}}^{\dag}(:,{\hat{\Omega}}_c^{l_1,k}){\tilde{\bf{Y}}}_k(:,{{\hat{\Omega}}_{r}}(l_1))$
\\7. \hspace*{+3mm} \textbf{end for}
\\8. \textbf{end for}
\\9. ${{\hat{\bf{H}}}_k}={{\bf{U}}_M^H}{{{\hat{\tilde{\bf{H}}}}}_k}{{\bf{U}}_N}, \forall k$
\\\textbf{Output}: Estimated cascaded channel matrices ${{\hat{{\bf{H}}}}}_k, \forall k$.
\end{algorithm}

The main procedure of \textbf{Algorithm 1} can be explained as follows. Firstly, the completely common row support $\Omega_r$ is jointly estimated thanks to the row-structured sparsity in Step 1, where $\Omega_r$ consists of $L_G$ row indexes associated with $L_G$ non-zero rows. Secondly, for the $l_1$th non-zero row, the partially common column support $\Omega_c^{l_1,{\rm{Com}}}$ can be further jointly estimated thanks to the partially column-structured sparsity in Step 2. Thirdly, the user-specific column supports for each user $k$ can be individually estimated in Step 3. After detecting supports of all sparse matrices, we adopt the LS algorithm to obtain corresponding estimated matrices $\{{{\hat{\tilde{\bf{H}}}}_{k}}\}_{k=1}^K$ in Steps 4-8. It should be noted that the sparse signal in~(\ref{eq8}) is ${\tilde{\bf{H}}}_{k}^{H}$, thus the sparse matrix estimated by the LS algorithm in Step 6 is ${\hat{\tilde{\bf{H}}}}_{k}^{H}$. Finally, we can obtain the estimated cascaded channels $\{{{\hat{{\bf{H}}}}_{k}}\}_{k=1}^K$ by transforming angular channels into spatial channels in Step 9.


In the following part, we will introduce how to estimate the completely common row support, the partially common column supports, and the individual column supports for the first three stages in detail.
\begin{algorithm}[htbp]
\caption{Joint completely common row support estimation}
\textbf{Input}: ${\tilde{\bf{Y}}}_k:\forall k$, ${L_G}$.
\\\textbf{Initialization}: ${\bf{g}}={\bf{0}}_{M\times 1}$.
\\1. \textbf{for} $k = 1,2,\cdots,K$ \textbf{do}
\\2. \hspace*{+3mm}${\bf{g}}(m)={\bf{g}}(m)+{\|{{\tilde{\bf{Y}}}_k}(:,m)\|}^2_F$, $\forall m=1,2,\cdots,M$
\\3. \textbf{end for}
\\4. ${\hat{\Omega}}_{r}={\Gamma}_{\mathcal{T}}({\bf{g}},L_G)$
\\\textbf{Output}: Estimated completely common row support ${\hat{\Omega}}_{r}$.
\end{algorithm}
\emph{1) Stage 1: Estimating the completely common row support.} Thanks to the row-structured sparsity of the angular cascaded channels, we can jointly estimate the completely common row support $\Omega_r$ for $\{{{{\tilde{\bf{H}}}}_{k}}\}_{k=1}^K$ by \textbf{Algorithm 2}.

From the virtual angular-domain channel representation~(\ref{eq4}), we can find that non-zero rows of $\{{{{\tilde{\bf{H}}}}_{k}}\}_{k=1}^K$ are corresponding to columns with high power in the received pilots $\{{{{\tilde{\bf{Y}}}}_{k}}\}_{k=1}^K$. Since $\{{{{\tilde{\bf{H}}}}_{k}}\}_{k=1}^K$ have the completely common non-zero rows, $\{{{{\tilde{\bf{Y}}}}_{k}}\}_{k=1}^K$ can be jointly utilized to estimate the completely common row support $\Omega_r$, which can resist the effect of noise. Specifically, we denote $\bf{g}$ of size $M\times 1$ to save the sum power of columns of $\{{{{\tilde{\bf{Y}}}}_{k}}\}_{k=1}^K$, as in Step 2 of \textbf{Algorithm 2}. Finally, $L_G$ indexes of elements with the largest amplitudes in $\bf{g}$ are selected as the estimated completely common row support ${\hat{\Omega}}_{r}$ in Step 4, where $\mathcal{T}({\bf{x}},L)$ denotes a prune operator on $\bf{x}$ that sets all but $L$ elements with the largest amplitudes to zero, and $\Gamma(\bf{x})$ denotes the support of ${\bf{x}}$, i.e., $\Gamma({\bf{x}})=\{i,{\bf{x}}(i)\neq0\}$.

After obtaining $L_G$ non-zero rows by \textbf{Algorithm 2}, we focus on estimating the column support ${{\Omega}}_c^{l_1,k}$ for each non-zero row $l_1$ and each user $k$ by the following Stage 2 and 3.

\emph{2) Stage 2: Estimating the partially common column supports.} Thanks to the partially column-structured sparsity of the angular cascaded channels, we can jointly estimate the partially common column supports $\{{{{\Omega}}_c^{l_1,{\rm{Com}}}}\}_{l_1=1}^{L}$ for $\{{{{\tilde{\bf{H}}}}_{k}}\}_{k=1}^K$ by \textbf{Algorithm 3}.
\begin{algorithm}[htbp]
\caption{Joint partially common column supports estimation}
\textbf{Input}: ${\tilde{\bf{Y}}}_k:\forall k$, $L_G$, ${\tilde{\bm{\Theta}}}$, ${L_{r,k}}:\forall k$, $L_c$, ${\hat{\Omega}}_{r}$.
\\\textbf{Initialization}: ${\hat{\Omega}}_{c}^{l_1,k}=\emptyset$, $\forall l_1,k$, ${\bf{c}}^l_1={\bf{0}}_{N\times 1}$, $\forall l_1$.
\\1. \textbf{for} $l_1 = 1,2,\cdots,L_G$ \textbf{do}
\\2. \hspace*{+3mm}\textbf{for} $k = 1,2,\cdots,K$ \textbf{do}
\\3. \hspace*{+6mm}${{\tilde{\bf{y}}}}_k={{\tilde{\bf{Y}}}}_k(:,{\hat{\Omega}}_{r}(l_1))$, ${{\tilde{\bf{r}}}}_k={\tilde{\bf{y}}}_k$
\\4. \hspace*{+6mm}\textbf{for} $l_2 = 1,2,\cdots,L_{r,k}$ \textbf{do}
\\5. \hspace*{+9mm}${n^{*}}={\mathop{\rm{argmax}}\limits_{n=1,2,\cdots,N}}{\|{\tilde{\bm{\Theta}}}^{H}(:,n){{\tilde{\bf{r}}}_k}\|}^2_F$
\\6. \hspace*{+9mm}${\hat{\Omega}}_{c}^{l_1,k}={\hat{\Omega}}_{c}^{l_1,k}\bigcup n^{*}$
\\7. \hspace*{+9mm}${\hat{\tilde{\bf{h}}}}_k={\bf{0}}_{N\times 1}$
\\8. \hspace*{+9mm}${\hat{\tilde{\bf{h}}}}_k({\hat{\Omega}}_{c}^{l_1,k})={\tilde{\bm{\Theta}}}^{\dag}(:,{\hat{\Omega}}_{c}^{l_1,k}){{\tilde{\bf{y}}}_k}$,
\\9. \hspace*{+9mm}${{\tilde{\bf{r}}}_k}={\tilde{\bf{y}}}_k-{\tilde{\bm{\Theta}}}{\hat{\tilde{\bf{h}}}}_k$
\\10.\hspace*{+9mm}${\bf{c}}^{l_1}(n^{*})={\bf{c}}^{l_1}(n^{*})+1$
\\11.\hspace*{+6mm}\textbf{end for}
\\12.\hspace*{+3mm}\textbf{end for}
\\13.\hspace*{+3mm}${\hat{\Omega}}_{c}^{l_1,{\rm{Com}}}=\Gamma_{{\mathcal{T}}({\bf{c}}^{l_1},P_c)}$
\\14. \textbf{end for}
\\\textbf{Output}: Estimated completely common row support $\{{\hat{\Omega}}_{c}^{l_1,{\rm{Com}}}\}_{l_1=1}^{L_G}$.
\end{algorithm}

For the $l_1$th non-zero row, we only need to utilize the effective measurement vector ${\tilde{\bf {y}}}_k={{\tilde{\bf{Y}}}}_k(:,{\hat{\Omega}}_{r}(l_1))$ to estimate the partially common column support ${\Omega}_c^{l_1,{\rm Com}}$. The basic idea is that, we firstly estimate the column support ${{\Omega}}_{c}^{l_1,k}$ with $L_{r,k}$ indexes for each user $k$, then we select $L_c$ indexes associated with the largest number of times from all $\{{{\Omega}}_{c}^{l_1,k}\}_{k=1}^{K}$ as the estimated partially common column support ${\hat{\Omega}}_{c}^{l_1,{\rm{Com}}}$.

In order to estimate the column supports for each user $k$, the correlation between the sensing matrix ${\tilde{\bm{\Theta}}}$ and the residual vector ${\tilde{\bf{r}}}_k$ needs to be calculated. As shown in Step 5 of \textbf{Algorithm 3}, the most correlative column index in ${\tilde{\bm{\Theta}}}$ with ${\tilde{\bf{r}}}_k$ is regarded as the newly found column support index $n^*$. Based on the updated column support ${\hat{\Omega}}_{c}^{l_1,k}$ in Step 6, the estimated sparse vector ${\hat{\tilde{\bf{h}}}}_k$ is obtained by using the LS algorithm in Step 8. Then, the residual vector ${\tilde{\bf{r}}}_k$ is updated by removing the effect of non-zero elements that have been estimated in Step 9. Particularly, the $N\times 1$ vector ${\bf{c}}^{l_1}$ is used to count the number of times for selected column indexes in Step 10. Finally, the $L_c$ indexes of elements with the largest value in ${\bf{c}}^{l_1}$ are selected as the estimated partially common column support ${\hat{\Omega}}_{c}^{l_1,{\rm{Com}}}$ in Step 13.
%

\emph{3) Stage 3: Estimating the individual column supports.} Based on the estimated completely common row support ${\hat{\Omega}}_{r}$ and the estimated partially common column supports $\{{\hat{\Omega}}_{c}^{l_1,{\rm{Com}}}\}_{l_1=1}^{L}$, the column support ${{\Omega}}_c^{l_1,k}$ for each non-zero row $l_1$ and each user $k$ can be estimated by \textbf{Algorithm 4}.
\begin{algorithm}[htbp]
\caption{Individual column supports estimation}
\textbf{Input}: ${\tilde{\bf{Y}}}_k:\forall k$, ${\tilde{\bm{\Theta}}}$, $L_G$ ${L_{r,k}}:\forall k$, $L_c$, ${\hat{\Omega}}_{r}$, $\{{\hat{\Omega}}_{c}^{l_1,{\rm{Com}}}\}_{l_1=1}^{L}$.
\\\textbf{Initialization}: ${\hat{\Omega}}_{c}^{l_1,k}={\hat{\Omega}}_{c}^{l_1,{\rm{Com}}}$, $\forall l_1,k$.
\\1. \textbf{for} $l_1 = 1,2,\cdots,L_G$ \textbf{do}
\\2. \hspace*{+3mm}\textbf{for} $k = 1,2,\cdots,K$ \textbf{do}
\\3. \hspace*{+6mm}${\tilde{\bf{y}}}_k={\tilde{\bf{Y}}}_k(:,{\hat{\Omega}}_{r}(l_1))$
\\4. \hspace*{+6mm}${\hat{\tilde{\bf{h}}}}_k={\bf{0}}_{N\times 1}$
\\5. \hspace*{+6mm}${\hat{\tilde{\bf{h}}}}_k({\hat{\Omega}}_{c}^{l_1,k})={\tilde{\bm{\Theta}}}^{\dag}(:,{\hat{\Omega}}_{c}^{l_1,{\rm{Com}}}){{\tilde{\bf{y}}}_k}$
\\6. \hspace*{+6mm}${\bf{r}}_k={\bf{y}}_k-{\tilde{\bm{\Theta}}}{\hat{\bf{h}}}_k$,
\\7. \hspace*{+6mm}\textbf{for} $l_2 = 1,2,\cdots,L_{r,k}-L_c$ \textbf{do}
\\8. \hspace*{+9mm}${n^{*}}={\mathop{\rm{argmax}}\limits_{n=1,2,\cdots,N}}{\|{\tilde{\bm{\Theta}}}^{H}(:,n){{\tilde{\bf{r}}}_k}\|}^2_F$
\\9. \hspace*{+9mm}${\hat{\Omega}}_{c}^{l_1,k}={\hat{\Omega}}_{c}^{l_1,k}\bigcup n^{*}$
\\10. \hspace*{+9mm}${\hat{\tilde{\bf{h}}}}_k={\bf{0}}_{N\times 1}$
\\11. \hspace*{+9mm}${\hat{\tilde{\bf{h}}}}_k({\hat{\Omega}}_{c}^{l_1,k})={\tilde{\bm{\Theta}}}^{\dag}(:,{\hat{\Omega}}_{c}^{l_1,k}){{\tilde{\bf{y}}}_k}$
\\12.\hspace*{+9mm}${{\tilde{\bf{r}}}_k}={\tilde{\bf{y}}}_k-{\tilde{\bm{\Theta}}}{\hat{\tilde{\bf{h}}}}_k$
\\13.\hspace*{+6mm}\textbf{end for}
\\14.\hspace*{+3mm}\textbf{end for}
\\15. \textbf{end for}
\\\textbf{Output}: Estimated the individual column supports $\{\{{\hat{\Omega}}_c^{l_1,k}\}_{l_1=1}^{L_G}\}_{k=1}^K$.
\end{algorithm}

For the $l_1$th non-zero row, we have estimated $L_c$ column support indexes by \textbf{Algorithm 3}. Thus, there are $L_{r,k}-L_c$ user-specific column support indexes to be estimated for each user $k$. The column support ${\hat\Omega}_{c}^{l_1,k}$ is initialized as ${\hat{\Omega}}_{c}^{l_1,{\rm Com}}$. Based on ${\hat{\Omega}}_{c}^{l_1,{\rm Com}}$, the estimated sparse vector ${\hat{\tilde{\bf{h}}}}_k$ and residual vector ${{\tilde{\bf{r}}}}_k$ are initialized in Step 5 and Step 6. Then, the column support ${\hat{\Omega}}_c^{l_1,k}$ for $\forall l_1$ and $\forall k$ can be estimated in Steps 7-13 by following the same idea of \textbf{Algorithm 3}.

Through the above three stages, the supports of all angular cascaded channels are estimated by exploiting the double-structured sparsity. It should be pointed out that, if there are no common scatters between the RIS and users, the double-structured sparse channel will be simplified as the row-structured sparse channel. In this case, the cascaded channel estimation can also be solved by the proposed DS-OMP algorithm, where Stage 2 will be removed.
\vspace{-3mm}
\subsection{Computational Complexity Analysis}
In this subsection, the computational complexity of the proposed DS-OMP algorithm is analyzed in terms of three stages of detecting supports. In Stage 1, the computational complexity mainly comes from Step 2 in \textbf{Algorithm 2}, which calculates the power of $M$ columns of ${\tilde{\bf{Y}}}_k$ of size $Q\times M$ for $k=1,2,\cdots,K$. The corresponding computational complexity is $\mathcal{O}(KMQ)$. In Stage 2, for each non-zero row $l_1$ and each user $k$ in \textbf{Algorithm 3} , the computational complexity $\mathcal{O}(NQL_{r,k}^3)$ is the same as that of OMP algorithm~\cite{XinyuGao_beamsqilt_TSP}. Considering $L_GK$ iterations, the overall computational complexity of \textbf{Algorithm 3} is $\mathcal{O}(L_GKNQ{L}_{r,k}^3)$. Similarly, the overall computational complexity of \textbf{Algorithm 4} is $\mathcal{O}(L_GKNQ{(L_{r,k}-L_c)}^3)$. Therefore, the overall computational complexity of proposed DS-OMP algorithm is $\mathcal{O}(KMQ)+\mathcal{O}(L_GKNQ{L}_{r,k}^3)$.

\section{Simulation Results}\label{S5}
In our simulation, we consider that the number of BS antennas, RIS elements and users are respectively $M=64$ ($M_{1}=8, M_{2}=8$), $N=256$ ($N_{1}=16$, $N_{2}=16$), and $K=16$. The number of paths between the RIS and the BS is $L_G=5$, and the number of paths from the $k$th user to the RIS is set as $L_{r,k}=8$ for $\forall k$. All spatial angles are assumed to be on the quantized grids. Each element of RIS reflecting matrix $\bm{\Theta}$ is selected from ${\{-\frac{1}{\sqrt{N}},+\frac{1}{\sqrt{N}}\}}$ by considering discrete phase shifts of the RIS~\cite{Wu'RIS}. $|{\alpha}^{G}_{l}|=10^{-3}d_{BR}^{-2.2}$, where $d_{BR}$ denotes the distance between the BS and RIS and is assumed to be $d_{BR}=10m$. $|{\alpha}^{r,k}_{l}|=10^{-3}d_{RU}^{-2.8}$, where $d_{RU}$ denotes the distance between the RIS and user and is assumed to be $d_{RU}=100m$ for $\forall k$~\cite{Wu'RIS}. The SNR is defined as $\mathbb{E}\{||{\tilde{\bm{\Theta}}}{\tilde{\bf{H}}}_{k}^{H}||_{F}^2/||{\tilde{\bf{W}}}_{k}||_{F}^{2}\}$ in (14) and is set as $0$ dB.

We compare the proposed DS-OMP based scheme with the conventional CS based scheme~\cite{JunCS} and the row-structured sparsity based scheme~\cite{LiangCS}. In the conventional CS based scheme, the OMP algorithm is used to estimate the sparse cascaded channel ${{\tilde{\bf{H}}}_k}$ for $\forall k$. In the row-structured sparsity based scheme, the common row support $\Omega_{r}$ with $L_G$ indexes are firstly estimated, and then for each user $k$ and each non-zero row $l_1$, column supports are respectively estimated by following the idea of the classical OMP algorithm. In addition, we consider the oracle LS scheme as our benchmark, where the supports of all sparse channels are assumed to be perfectly known.

\begin{figure}[htpb]
\begin{center}
\includegraphics[width=0.8\linewidth]{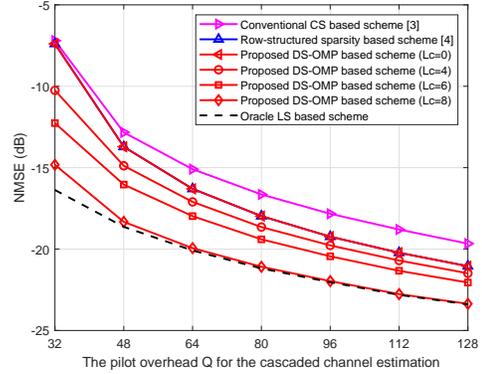}
\end{center}
\setlength{\abovecaptionskip}{-0.0cm}
\caption{NMSE performance comparison against the pilot overhead $Q$.} \label{FIG4}
\vspace{-4mm}
\end{figure}

Fig. 2 shows the normalized mean square error (NMSE) performance comparison against the pilot overhead, i.e., the number of time slots $Q$ for pilot transmission. As shown in Fig. 2, in order to achieve the same estimation accuracy, the pilot overhead required by the proposed DS-OMP based scheme is lower than the other two existing schemes~\cite{JunCS, LiangCS}. However, when there is no common path between the RIS and all users, i.e., $L_c=0$, the double-structured sparsity will be simplified as the row-structured sparsity~\cite{LiangCS}. Thus the NMSE performance of the proposed DS-OMP based and the row-structured sparsity based scheme is the same. With the increased number of common paths $L_c$ between the RIS and users, the NMSE performance of the proposed scheme can be improved to approach the benchmark of perfect channel supports.
\vspace{-2mm}

\section{Conclusions}\label{S6}
In this paper, we developed a low-overhead cascaded channel estimation scheme in RIS assisted wireless communication systems. Specifically, we first analyzed the double-structured sparsity of the angular cascaded channels among users. Based on this double-structured sparsity, we then proposed a DS-OMP algorithm to reduce the pilot overhead. Simulation results show that the pilot overhead required by the proposed DS-OMP algorithm is lower compared with existing algorithms. For the future work, we will apply the double-structured sparsity to the super-resolution channel estimation problem by considering the channel angles are continuous in practice.

\bibliography{IEEEabrv,Part_II}

\end{document}